\documentclass[aps,twocolumn,prd,epsf,nofootinbib]{revtex4}

\usepackage{amssymb}
\usepackage{graphicx}
\usepackage{soul}
\usepackage{amsmath}
\usepackage{amsfonts}
\usepackage{comment} 
\usepackage{placeins}
\usepackage{float}
\usepackage{mathrsfs}
\newcommand{\be}{\begin{equation}}
\newcommand{\ee}{\end{equation}}
\newcommand{\bea}{\begin{eqnarray}}
\newcommand{\eea}{\end{eqnarray}}
\newcommand{\mpl}{m_{\rm pl}}

\newcommand{\di}{ {\rm d}}
\usepackage{color}

\begin{document}

\title{Dissecting the growth of the power spectrum for primordial black holes}
\author{Pedro Carrilho}
\email{p.gregoriocarrilho@qmul.ac.uk}
\author{Karim A. Malik}
\email{k.malik@qmul.ac.uk}
\author{David J. Mulryne }
\email{d.mulryne@qmul.ac.uk}
\affiliation{\vspace{2mm}
School of Physics and Astronomy, Queen Mary University of London, Mile End Road, London, E1 4NS, UK}

\date{\today}
\begin{abstract}
We consider the steepest rate at which the power spectrum from single
field inflation can grow, with the aim of providing a simple
explanation for the $k^4$ growth found recently.  With this
explanation in hand we show that a slightly steeper $k^5 (\log k )^2$
growth is in fact possible.  Moreover, we argue that the power
spectrum after a steep growth cannot immediately decay, but must
remain large for the $k$ modes which exit during a $\sim2$ e-fold
period.  We also briefly consider how a strong growth can affect the
spectral index of longer wavelengths preceding the growth, and show
that even the conversion of isocurvature modes likely cannot lead to a
stronger growth.  These results have implications for the formation of
primordial black holes, and other phenomena which require a large
amplitude of power spectrum at short scales.

\end{abstract}
\maketitle

\section{Introduction}

How steeply can ${\cal P}_{\cal R}$, the power spectrum of the primordial
curvature perturbation, grow with scale? This interesting question is
addressed in the recent paper of Byrnes {\it et al.}
\cite{ByrnesEtAl}.  If produced during inflation, one might think that
the classic relation, ${\cal P}_{\cal R}\sim
{H_*^2}/{\epsilon_*}$, implies that a model with a sudden
change in $\epsilon$ to a much smaller value would allow an
arbitrarily steep growth, through for example the rapid flattening of
the inflationary potential.  Here $H_*$ is the Hubble rate at which
  a given scale is equal to the cosmological horizon during inflation,
  and $\epsilon_*$ is the value of slow-roll parameter $\epsilon = -
  \dot{H}/H^2$ at that time.  However, the slow-roll solution to the
equations of motion that the above relation for ${\cal P}_{\cal R}$ 
relies on becomes invalid for
a period around such a flattening. Byrnes {\it et al.}\cite{ByrnesEtAl} therefore
explored analytic solutions to the evolution of perturbations in
models with $\epsilon \ll 1$ (so inflation always proceeds), but with
the second slow-roll parameter, $\eta= \dot \epsilon/(H \epsilon)$, 
taking a series of discrete
values between $0$ and large negative values, matching the
perturbations and their first derivative at each change in
$\eta$. This setup encompasses, but is more general than, 
the case of a rapid flattening of the inflationary potential 
described above. 
Doing so, the steepest sustained growth that they observed was
in fact proportional to $k^4$, which they then argued is the steepest
possible.  Numerical simulations backed up their conclusion.

This result is of considerable interest. The discovery by LIGO of
binary in-spiralling black holes with masses of the order of a few
$M_\odot$ \cite{Abbott:2016blz,LIGOScientific:2018jsj,LIGOScientific:2018mvr}, as well as the continued absence of a microphysical
explanation for dark matter, have lead to a renewed interest in
primordial black holes (PBHs) \cite{Carr:1974nx,Bird:2016dcv,Sasaki:2016jop,Carr:2016drx,Carr:2017jsz}.  
If PBHs form in
the radiation era which follows inflation, however, ${\cal P}_{\cal R}$ must grow
by at least seven orders of magnitude between the scales which account
for the formation of large scale structure in the universe and the
scales which lead to PBHs \cite{Zaballa:2006kh,Josan:2009qn,Cole:2017gle,Germani:2017bcs,Germani:2018jgr}.  Moreover, if PBHs are to
form only in specific mass ranges to evade constraints \cite{Carr:2009jm,Carr:2017jsz,ByrnesEtAl}, the particular 
form of the growth and decay of the power spectrum is very
important. The result of Ref.~\cite{ByrnesEtAl} is also of interest
more generally, for example a large power spectrum at short scales can
produce gravitational waves at second order in perturbation theory \cite{Ananda:2006af,Baumann:2007zm,Alabidi:2012ex,Gong:2017qlj,Kohri:2018awv,Clesse:2018ogk}.

In this short paper we build on the work 
of Byrnes  {\it et al.} \cite{ByrnesEtAl} 
by studying
the same system. Our aim is to 
provide a clear and simple
explanation of where the $k^4$ growth comes from, and with this
explanation in hand to ask whether $k^4$ growth really is the steepest
sustained growth in $k$ space that is possible. We show that this is in fact not
the case, and that it is possible to have a steeper sustained 
growth proportional to $k^5(\log k )^2$ in single field inflation.  We also consider how the spectrum 
preceding the growth is affected by the behaviour needed to produce a strong growth, and
comment on the subsequent rate of decay which cannot be instantaneous. Finally we apply the 
insight we develop to the isocurvature sector to see if it could in principle
be responsible for an even steeper growth of ${\cal P}_{\cal R}$ -- we find this is likely
not possible.  

\section{slow-roll to ultra-slow-roll}
\label{SR2USR}

\begin{figure}[]
\centerline{\includegraphics[angle=0,width=49mm,height=36mm]{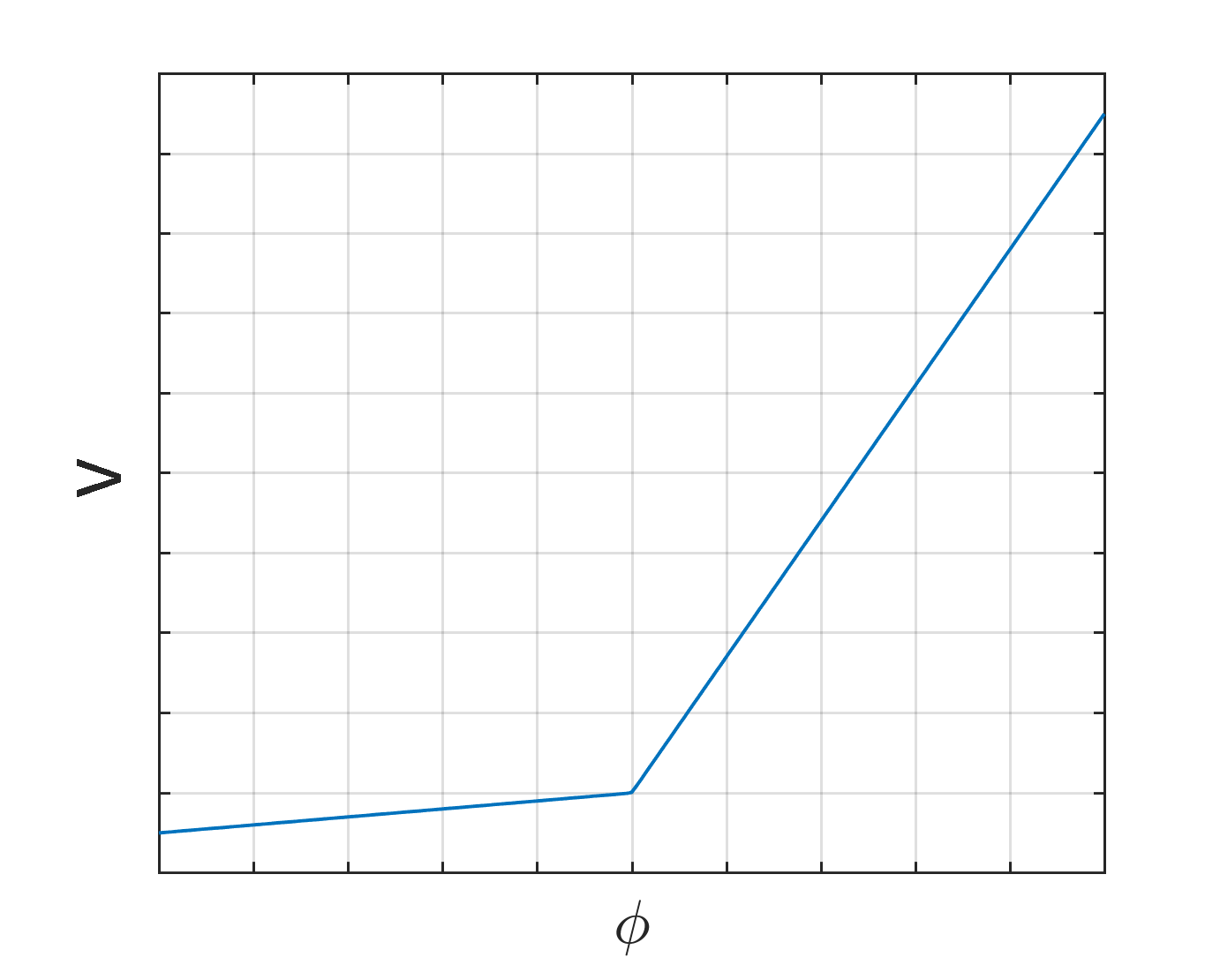} \hspace{-.4cm} \includegraphics[angle=0,width=49mm]{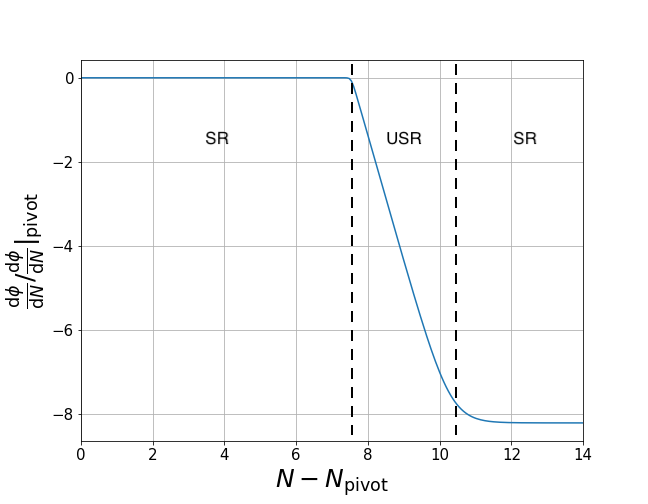}} 
\centerline{\includegraphics[angle=0,width=49mm]{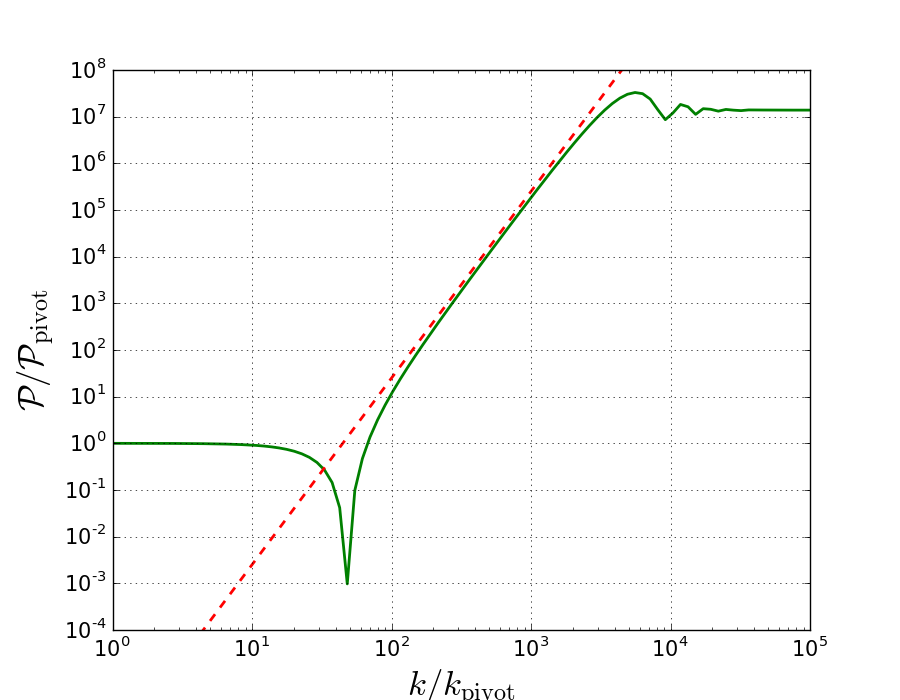}\hspace{-.4cm} \includegraphics[angle=0,width=49mm]{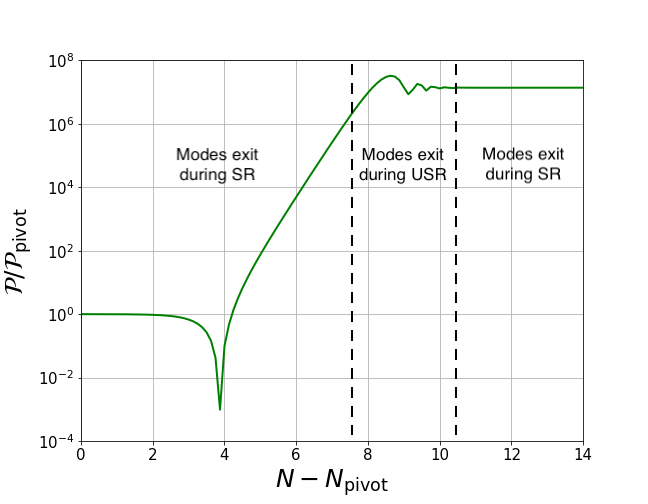}} 
\caption{a (top left) shows schematically a potential that rapidly flattens; b (top right) shows the rate the field changes as it rolls along this potential, the $a^{-3}$ scaling can 
clearly been seen; c (bottom left)  shows the resultant power spectrum against $k$, with $k^4$ dependence 
indicated by the dashed line;  and d (bottom right) shows the power spectrum against exit time in e-folds, highlighting the regime in 
which different modes exit.} 
\label{PyTPlot}
\vspace*{-15pt}
\end{figure}

Let us follow Byrnes {\it et al.} \cite{ByrnesEtAl}, and others \cite{Germani:2017bcs,Ballesteros:2017fsr,Gong:2017qlj,Cheng:2018qof,Bhaumik:2019tvl}, 
and study first the most natural situation which leads to a steep growth 
in the power spectrum. Consider 
canonical single-field inflation with a potential $V(\phi)$ with small
gradient $V'_1$ with $V'_1\ll V$, followed by a sudden transition to
gradient $V'_2 \ll V'_1$, as illustrated in Fig.~(\ref{PyTPlot}a).
Fourier modes of the comoving curvature perturbation ${\cal R}$\footnote{See, for example, Ref.~\cite{Malik:2008im} for a full definition} 
which are able to transition between their
vacuum state and their asymptotic form, in which they are constant,
while slow-roll inflation occurs on the $V'_1$ region have a close to
scale-invariant spectrum: ${\cal P}_{\cal R}(k) \sim {V^4}/{V'_1}^2$,
while modes which can move from vacuum to asymptotic form on the
$V'_2$ region have spectrum ${\cal P}_{\cal R}(k) \sim {V^4}/{V'_2}^2$.
The velocity of the inflaton field, $\di \phi / \di N$, for the slow
roll attractor solution in the two regimes is $-V'_1/V$ and $-V'_2/V$,
respectively ($N\equiv\log a$ is the number of e-folds, $a$ the scale factor).  This implies
that at the transition the velocity is initially much higher than the
second slow-roll value, and must decay to reach it. Initially the
gradient of the potential is irrelevant, and the steepest that this can
happen is at the rate for a free field of $\di \phi/\di N \propto
a^{-3}$, where we assume the Hubble rate is approximately constant.
This represents a phase of ultra-slow-roll (USR) inflation, followed
by slow-roll inflation again at the new lower velocity, as can be
seen in Fig.~(\ref{PyTPlot}b). This behaviour leads to an approximately
scale-invariant spectrum for modes which exit sufficiently early in
the first slow-roll phase followed by a spectrum which grows in
proportion to $k^4$, and finally a near scale-invariant spectrum
again, as seen in Figs.~(\ref{PyTPlot}c) and (\ref{PyTPlot}d) where
we indicate the phases during which perturbations exit the horizon. We
have generated these figures using the open source code {\it
  PyTransport} \cite{Dias:2016rjq,Mulryne:2016mzv,Ronayne:2017qzn} (see also \cite{Dias:2016rjq,Seery:2016lko,Butchers:2018hds} for a related package).

The question at hand now is what is the origin of the $k^4$
dependence. As we
can see from Fig.~(\ref{PyTPlot}) modes with the $k^4$ dependence exit
during slow-roll inflation. We therefore consider the Sasaki-Mukhanov
equation
\be
v''_k + \left(k^2 - \frac{z''}{z}\right) v_k =0\,,
\label{eqSM}
\ee 
where a dash indicates differentiation with respect to conformal time, $\tau$, and where
\be
\label{Defz2}
z^2=2a^2\mpl^2 \epsilon\,,
\ee
and
\be
\frac{z''}{z} = (a H^2) \left (2 -\epsilon + \frac{3}{2}\eta + \frac{1}{4}\eta^2 - \frac{1}{2} \epsilon \eta +\frac{1}{2} \frac{\dot \eta}{H}\right)\,,
\ee
where a dot indicates differentiation with respect to coordinate time.
Recall also that $a H = -1/\tau$, and
that $v$ is related to ${\cal R}$ through ${\cal R}= v/z$.  Assuming $\epsilon\ll1$ and defining
\be
\nu^2 = \frac{9}{4} + \frac{3}{2} \eta + \frac{1}{4} \eta^2
+\frac{\dot \eta}{2 H}\,,
\ee
one finds that for constant $\nu$ Eq.~(\ref{eqSM}) has the general
solution
\be v_k = \tilde A
\sqrt{-\tau} H_\nu^{(1)}(- k \tau) + \tilde B \sqrt{-\tau}
H_\nu^{(2)}(- k \tau) \,.
\label{general}
\ee
Note that for constant $\eta$, $\epsilon \propto
(-\tau)^{-\eta}$. For modes which begin in the Bunch-Davies vacuum
the solution is then normalised by fixing the constants $\tilde A$ and $\tilde
B$ such that
\be v_k = \frac{\sqrt{\pi}}{2}
e^{i \frac{\pi}{2} \left( \nu+\frac{1}{2}\right)} \sqrt{-\tau}H_\nu^{(1)}(- k
\tau)\,.
\label{bunch}
\ee
Expanding in the super-horizon limit, $k \tau \to 0$, one finds
\begin{eqnarray}
v_k  &\approx& \frac{\sqrt{\pi}}{2} e^{i \frac{\pi}{2} \left( \nu+\frac{1}{2}\right)} \sqrt{-\tau} \left ( -2^{\nu}i\frac{\Gamma(\nu)}{\pi}(-k \tau )^{-\nu} \right . \nonumber \\ 
 &~&- 2^{-2+\nu}i\frac{\Gamma(\nu)}{\pi(-1+\nu)}(-k \tau )^{-\nu+2} \nonumber \\
&~& \left .
 2^{-\nu}\frac{1+i\cot(\pi \nu) }{\Gamma(2+\nu)}(-k \tau )^{\nu}    + \dots
\right)\,,
\label{genExp}
\end{eqnarray}
where we have assumed $\nu$ is positive, but that $\nu \neq 1$. For ${\cal R}$ this implies
\begin{eqnarray}
{\cal R}_k &\approx& A\, (-\tau)^{-\nu + \frac{3}{2} + \frac{\eta}{2}} k^{-\nu} + B\, (-\tau)^{-\nu + \frac{7}{2} + \frac{\eta}{2}}k^{-\nu+2} \nonumber \\ &+&  C\, (-\tau)^{\nu + \frac{3}{2} + \frac{\eta}{2}}k^{\nu} \,,
\label{genExpZ}
\end{eqnarray}
where in this equation we note that $A$, $B$ and $C$ are constants
that are not free but take fixed values.  Here we have retained three
terms because it is not clear which of the last terms is the
sub-leading term for a general $\nu$.  Finally considering modes which
exit during slow-roll inflation, one has $\eta \to0$, $\nu \to 3/2$, and in this
case one finds to sub-leading order that
\be
{\cal R}_k \approx A  k^{-3/2} + B (-\tau)^{2} k^{1/2 }\,.
\label{InfExp}
\ee
Given that ${\cal P}_{\cal R}(k) \sim k^3 |{\cal R}|^2$, one can see that if
the leading (constant) term dominates (as is usually expected), ${\cal
  P}_{\cal R}\propto k^0$, while if the sub-leading term would dominate
the resulting scale-dependence would be ${\cal P}_{\cal R}\propto k^4$.

\subsection{Matching}

Now let us consider what happens if there is a transition to the USR
phase. For this phase $\nu =-3/2$ and $\eta =-6$ and on scales larger 
than the horizon the general solution (\ref{general}) gives
\be
{\cal R}_k =A_2 (-\tau)^{ -3}  +  B_2 \dots\,,
\label{afterUSR1}
\ee
where the subscript $2$ indicates quantities after the transition, and
where we have only kept the leading term and sub-leading terms with
independent coefficients $A_2$ and $B_2$. Note that one of 
the independent solutions to Eq.~(\ref{eqSM}) for any $\nu$ will always lead to 
a constant ${\cal R}$ on super-horizon scales.
 To understand the fate of
modes which exited during slow-roll inflation we now match solution
(\ref{InfExp}) to (\ref{afterUSR1}).  Matching both the
solution and its first derivative, as required by the Israel junction
conditions \cite{Israel:1966rt,PhysRevD.52.5549}, at the time $\tau=-1$ (which is an arbitrary but
convenient choice)\footnote{This choice also acts to fix the units being used in this and other calculations, such that all conformal times are measured with respect to the transition time.} we find
\begin{eqnarray}
A k^{-3/2} &\approx&  A_2 + B_2\,,\\
2 B k^{1/2}  &\approx& -3 A_2 \,,
\end{eqnarray}
and hence  $A_2 \approx -2/3 B k^{1/2}$,  $B_2 \approx A k^{-3/2}$, and
\be
{\cal R}_k =-2/3 B k^{1/2} (-\tau)^{ -3}  +  A k^{-3/2}+\dots\,.
\label{afterSRUSR2}
\ee
Moreover, recall that $A$ and $B$ are fixed by the Bunch-Davies origin
of the modes, and one can verify that $A \sim B$.  
 Since we are
considering modes with $k\ll 1$, at the matching time the second term
in Eq.~(\ref{afterSRUSR2}), which would lead to a scale-invariant
spectrum, is initially the dominant one. However the second term, which has inherited the 
scale dependence of the decaying mode before the transition, is
growing. For a given mode, therefore, if USR lasts sufficiently long this
term will become dominant. If this happens for a range of scales, the
spectrum for these modes gains a $k^4$ dependence, explaining the
origin of this dependence. As was 
also discussed by Byrnes {\it et  al.}\cite{ByrnesEtAl}, the duration 
  of the USR phase therefore plays a crucial role.
The ratio of the power spectrum before and after the steep growth, $P_{{\cal R}\, \text{ratio}}$, is approximately given 
by the square of the ratio of the two terms in 
Eq.~(\ref{afterSRUSR2}). Defining  $\tau_{\text{USR}} =  \tau_{\text{USR\,start}}/\tau_{\text{USR\,end}} $, therefore, we
see that $P_{{\cal R}\, \text{ratio}} \approx  \tau_{\text{USR}}^{6} $. 
For $P_{{\cal R}\, \text{ratio} }= 10^7$, needed for PBH formation, then implies
USR must last $\log 10^{7/6} \approx 2.7$ e-folds. Equation~(\ref{afterSRUSR2}) also explains the dip 
in the power spectrum  seen in Fig.~(\ref{PyTPlot}c) which precedes the $k^4$ dependence.    
This
occurs because the two terms in Eq.~(\ref{afterSRUSR2}) approximately cancel
one another, and so for a given scale preceding the scales which show
a clear $k^4$ dependence, they in fact cancel each other exactly
(at least up to the effect of other sub-dominant terms).  
Labelling $k_{\rm peak}$ as the mode which exits the horizon just as 
USR starts, Eq.~(\ref{afterSRUSR2}) gives $k_{\rm dip} = \tau_{\text{USR}}^{-3/2} k_{\rm peak}$. For $P_{{\cal R}\, \text{ratio}} =10^{7}$ this 
gives $\log_{10}{(k_{\rm dip}/k_{\rm peak})} \approx  -7/4 $. This cancellation was already noticed in Ref.~\cite{ByrnesEtAl}, as well as having been studied in Ref.~\cite{Passaglia:2018ixg}, and found to have interesting consequences for non-Gaussianity generated in these scenarios.

We note that Byrnes {\it et  al.}\cite{ByrnesEtAl} arrived at a similar expansion to (\ref{afterSRUSR2}) 
 above after matching the full unexpanded solution  (\ref{general}) before and after the transitions,
 and subsequently expanding.
{\it The crucial new point to 
appreciate in our analysis, however,  is that the origin of the
scale-dependence of the growing term is that of the sub-leading
(decaying) term in an expansion before the matching.}

\subsection{Lesson}
\label{lesson}

If the leading term in
the expansion of ${\cal R}$ before the transition were not constant, only
the leading term would be imprinted on the solution after the
transition. In this case the scale-dependence of the leading term
would always be transmitted through the transition, and the final
spectrum of perturbations would be given by
\be
n_{\rm s} - 1 = 3 - | 3+\eta |\,,
\ee
and hence be bounded according to the condition $ n_{\rm s} - 1 <
3$.  

If the leading term is constant, however, and 
there is a growing mode after the transition, the scale-dependence of the sub-leading term
before the transition is inherited by this growing mode. Here we have
considered a sudden change to a phase with a growing mode, but the
same conclusion must hold even if the transition is more gradual. {\it
  The allowed range of scale-dependence can therefore be determined by
  considering both the leading and sub-leading terms} in the solution
to the Sasaki-Mukhanov equation normalised to a Bunch-Davies vacuum.

\section{A steeper spectrum}
\label{steeper}

What then is the bound on 
the steepness of the power spectrum in cases where the leading term before the transition 
is constant?
Considering Eq.~(\ref{genExpZ}), one finds that  $\eta > -3$ leads to a constant ${\cal R}$. 
We can therefore answer this question by looking exhaustively 
at these cases and the scale-dependence of the sub-leading term for each.

Considering again Eq.~(\ref{genExpZ}) we see that the sub-dominant
term is the $B$ term for $\eta>-1$ and the $C$ term for $\eta<-1$. For
this reason, the scale-dependence $k^\alpha$ of the sub-leading term
is given by
\begin{eqnarray}
\label{specindfromdecay}
\alpha=1-\frac12 |\eta+1|\,,
\end{eqnarray}
which peaks in the case $\eta \to -1$. But $\eta = -1$, which gives $\nu = 1$, is a special case in which the expansion as given above 
breaks down, and the sub-leading term as written blows up (and hence ceases to be sub-leading when that 
limit is approached). In fact for $\eta =-1$, the 
expansion of solution (\ref{bunch}), leads to
\begin{eqnarray}
{\cal R}_k &\approx& A k^{-1} + B (-\tau ) (-k \tau) ( D+\log(-k \tau))\,,\label{better}\\
 &\approx& A k^{-1} + B (-\tau )^2 k\log(-k \tau) \,, \label{subMax}
\end{eqnarray}
before the transition, where $D =\gamma-1/2-\log2-\pi i$, in which $\gamma\approx 0.577216$ is Euler's constant. We drop the constant term $D$ in the second line 
above for simplicity, but note that it only becomes negligible slowly in the $-k\tau \to 0$ limit, 
and a better approximate scaling (used in Fig.~(\ref{1toUSR})) comes from retaining this term.
An $\eta =-1$ phase therefore gives the steepest scaling of the
sub-leading term that can be realised in the asymptotic limit.  Considering modes which exit
the horizon during this phase, and matching to a phase after the
transition in which there is a growing mode (for example an USR phase
-- Eq.~(\ref{afterUSR1})), we find
\begin{eqnarray}
A k^{-1} &\approx&  A_2 + B_2 \,,\\
2 B k \log k   &\approx& -3 A_2 \,,
\end{eqnarray}
and hence $A_2 \approx -2/3 B k \log k $, $B_2 \approx A k^{-1}$ and so
after the transition in this case we get
\be
{\cal R}_k =-2/3 B k \log k  (-\tau)^{ -3}  +  A k^{-1}+\dots\,.
\label{afterSRUSR}
\ee
As the first term grows relative to the second term, the resulting
spectrum is of the form ${\cal P}_{\cal R}\propto k^5
(\log k) ^2$. Recalling that we have matched at time $\tau=- 1$, we see
that modes we are considering satisfy $k\ll1$, and hence this form of
spectrum sits somewhere between growth proportional to $k^4$ and
$k^5$, but \emph{is} steeper than $k^4$, with $k^5$ being the limiting value which can never 
quite be reached.

\subsection{Example spectra}

\begin{figure}[]
\centerline{ \hspace{.1cm}\includegraphics[angle=0,width=49mm]{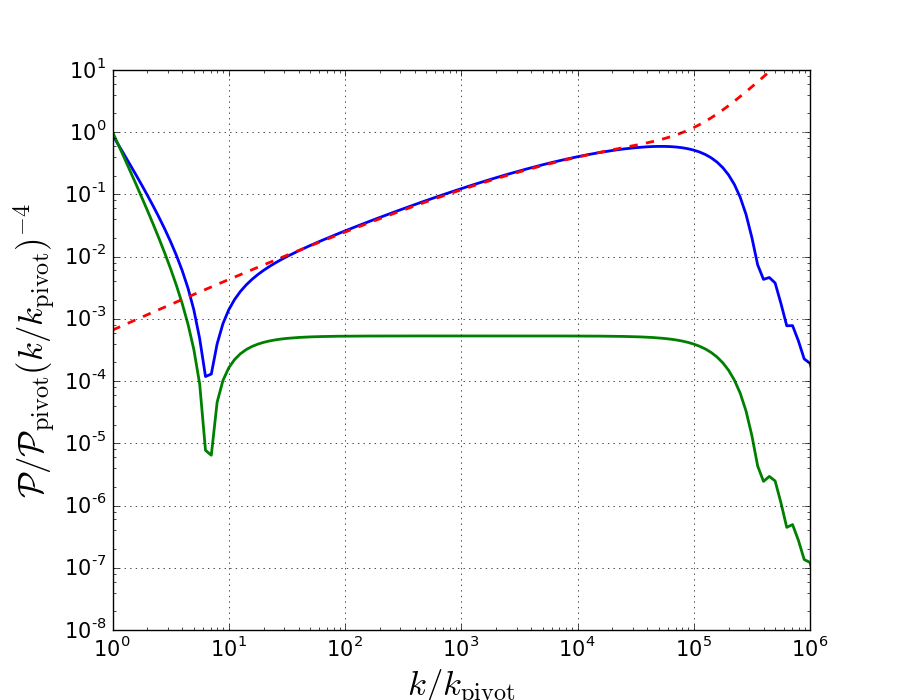} \hspace{-.5cm} \includegraphics[angle=0,width=49mm]{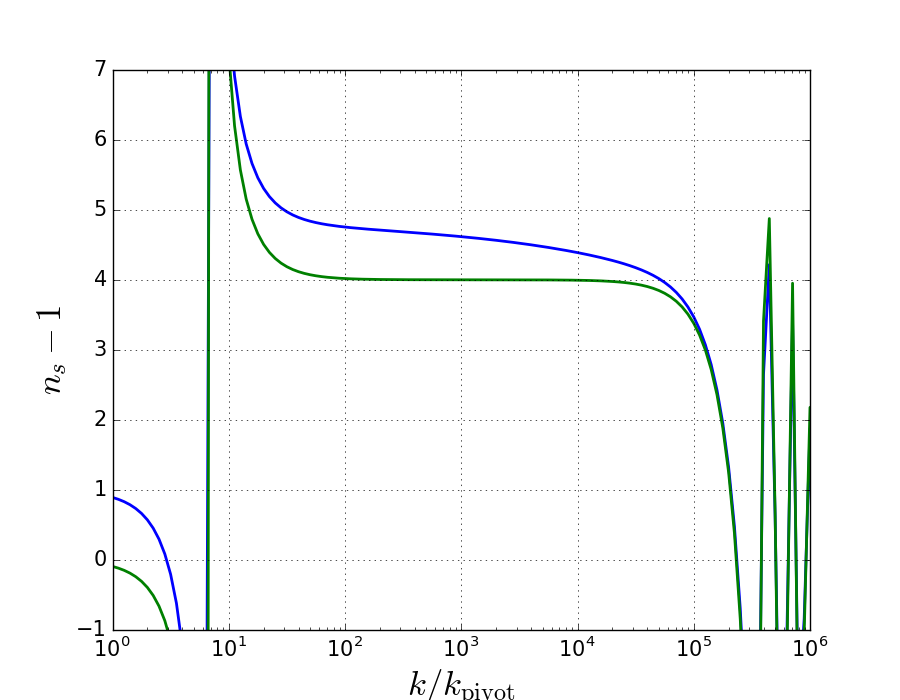}} 
\centerline{ \hspace{.1cm}\includegraphics[angle=0,width=49mm]{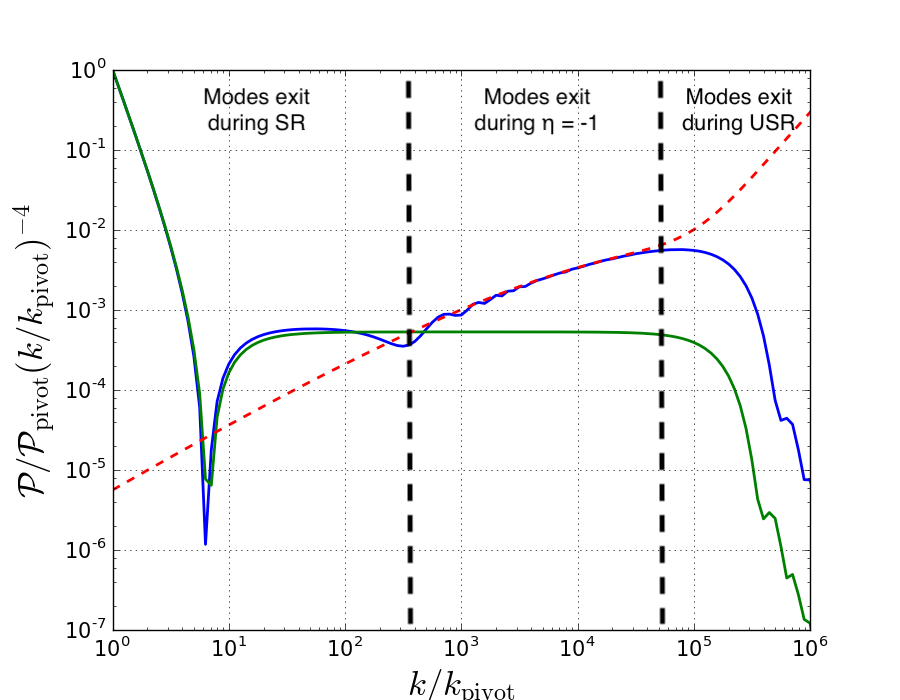}\hspace{-.5cm} \includegraphics[angle=0,width=49mm]{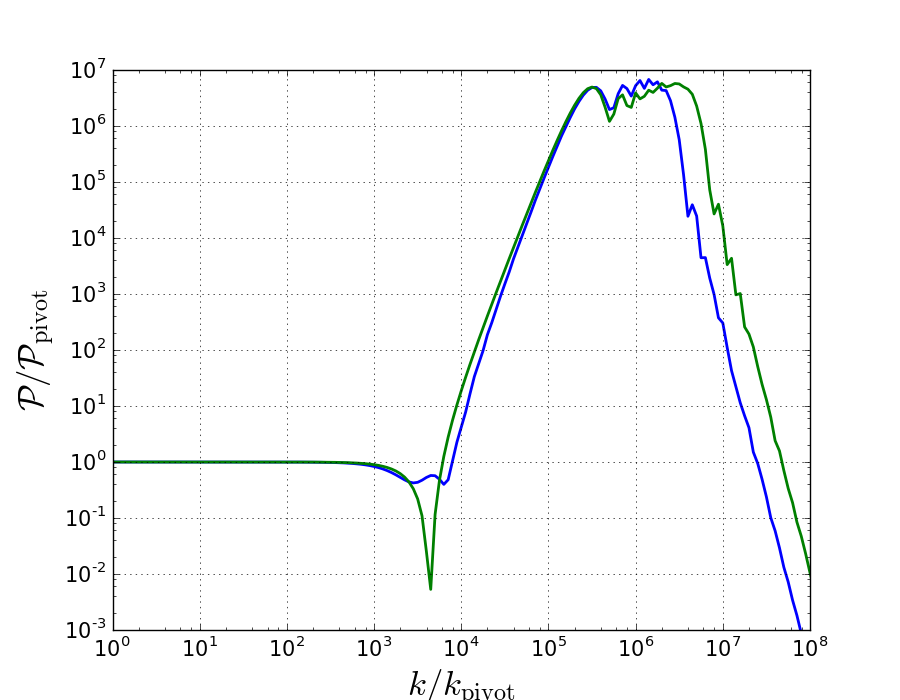}} 
\caption{a  (top left) compares the power spectra scaled by $k^4$ for the transitions $\eta=0\rightarrow\eta=-6$ (green) and $\eta=-1\rightarrow\eta=-6$ (blue), showing the steeper growth possible for the latter case.  The duration of the USR stage was chosen to make the dip in the two spectra match. The figure also shows the $k^5|(0.1+1.6i) -  \log k|^2$ scale dependence  (red dashed line), which comes from using Eq.~(\ref{better}) to do the matching 
(see text below Eq.~(\ref{better})); b (top right) shows the spectral index for these cases; c (bottom left) 
compares the power spectra scaled by $k^4$ for the transitions $\eta=0\rightarrow\eta=-6$ (green) and $\eta=0\rightarrow\eta=-1\rightarrow\eta=-6$ (blue), together with the same approximate scaling as 
above (red dashed). The dashed lines demarcate the phase in which modes exit for the   $\eta=0\rightarrow\eta=-1\rightarrow\eta=-6$ case; d (bottom right)  compares the power spectra for the transitions $\eta=0\rightarrow\eta=-6\rightarrow\eta=6$ (green) and $\eta=0\rightarrow\eta=-1\rightarrow\eta=-6\rightarrow\eta=6$ (blue), showing the shorter range of scales exiting the horizon during the USR stage in the latter case.}
\label{1toUSR}
\vspace*{-15pt}
\end{figure}

The $\eta = -1$ behaviour can be achieved numerically in principle, using the potential reconstruction derived in Appendix C of Ref.~\cite{ByrnesEtAl}.\footnote{For constant $\eta$, this potential is given by 
\begin{equation}
V(\phi)=V_*\exp{\left[-\frac{\eta+6}{24}\eta \frac{(\phi-\phi_*)^2}{\mpl^2}-\frac{\eta+6}{6}\sqrt{2\epsilon_*}\frac{\phi-\phi_*}{\mpl}\right]}\,,
\end{equation}
in which stared variables are evaluated at the start of the current phase.} However, for simplicity we follow here the the analytic approach of Ref.~\cite{ByrnesEtAl} to provide examples of the full resulting spectra. We repeat the matching 
procedure described above, but match the full solutions 
for ${\cal R}$  and ${\cal R}_k'$ (which follow from Eq.~\ref{general}) 
before and after the change in the value of $\eta$.
This results in a solution after the transition that is valid on all scales.
For the case of a $\eta=-1 \rightarrow \eta=-6$ transition, we plot the power spectrum in Fig.~(\ref{1toUSR}a), and $n_{\rm s}$  in Fig.~(\ref{1toUSR}b).

We can clearly see that the spectral index is larger than $4$ for this
new transition and that it matches the approximate 
$k^5(\log k )^2$ form predicted above. This demonstrates again the 
point that the scaling of the steep growth is determined by the 
$k$ dependence of the sub-leading term in an expansion for modes exiting before 
the USR stage.
We also emphasise that although 
not of power law form, this behaviour is 
sustained over a large range of scales  (i.e. not a transitory growth 
caused, for example, by the cancellation of two terms in an asymptotic expansion, 
as occurs after the dip)
just like the $k^4$ growth.
This is the steepest sustained growth we could find in all cases tested.  We have 
verified that the scale dependence does not change for transitions into stages
other than USR (i.e.~$\eta=-5$ or $\eta=-4$), as was also the case for
the results of Ref.~\cite{ByrnesEtAl}, which we again stress is
expected since the form of the spectrum is determined by the modes exiting before
the transition. Finally we investigated examples with many different $\eta$ values in the pre-transition 
phase. In all cases
the scale dependence of the spectra matches the expectations of the argument of section \ref{lesson}.

It is well known that for scales that are observed in the Cosmic Microwave Background (CMB) the
spectrum must be approximately scale-invariant and hence a stage with
$\eta\approx 0$ is always required.
To present a more realistic example of a growth steeper than $k^4$
therefore, we considered multiple transitions, such as
$\eta=0\rightarrow\eta=-1\rightarrow\eta=-6$. In this case one still expects
modes which exit the horizon during the $\eta = -1$ phase to have a
steeper than $k^4$ dependence assuming  USR lasts long enough. We
tested this scenario by having a  $\eta=0\rightarrow\eta=-1$ transition 
at $\tau_{1,2}=-500$ and a second $\eta=-1\rightarrow\eta=-6$ transition
 at $\tau_{2,3}=-1$. We show the results in
Fig.~(\ref{1toUSR}c).
For the parameters chosen, the scales exiting the
horizon during the $\eta=-1$ stage are all enhanced by the subsequent
USR stage giving a $k^5(\log k)^2$ type scaling, while the final modes that crossed
the horizon during the slow-roll stage ($\eta=0$) are also enhanced but give rise to
a $k^4$ scaling. 
Here the duration of the USR phase is once again 
extremely important. The approximate dependence of the scale of the dip,
$k_{\text{dip}}$, with the duration of the USR stage remains 
$k_{\rm dip} \approx \tau_{\text{USR}}^{-3/2} k_{\rm peak}$.  This is because the time dependence of the 
decaying mode during $\eta=-1$ differs only logarithmically when compared with the slow-roll phase, and 
at a slow-roll to $\eta=-1$ transition the decaying mode in slow-roll is 
matched  to decaying mode of the $\eta=-1$ phase. Assuming USR is long enough 
for $k_{\rm dip}$ to correspond to scales which crossed the  horizon during slow-roll, 
as in Fig.~(\ref{1toUSR}c),
the $k^4$ dependence is seen for slightly shorter scales, followed by a transition to $k^5(\log k)^2$ 
type scaling at $k_{\rm trans} \approx \tau_{1,2}^{-1}$.

\section{Decay of the spectrum}
\label{decay}

A key difference between the slow-roll to USR transition and the case with 
an intermediate $\eta=-1$ phase, is that the steeper growth that 
this phase allows means that the USR stage can be shorter for the 
same overall growth. In Fig.~(\ref{1toUSR}d), we illustrate this point with an
example where the parameters have been chosen to maximise the range of
scales affected by the $k^5(\log k )^2$ spectrum.  In this case, the
duration of the USR is now $\approx 1.65$ e-folds
compared with $\approx 2.3$ e-folds for the pure slow-roll to USR case 
(calculated using the full matching).

We now stress a second important consequence of the essential USR stage.
During USR, modes still exit the horizon with 
an approximately scale-invariant spectrum. 
After the spectrum
has reached its maximum, therefore, a non-negligible range of scales must be
approximately scale-invariant and the spectrum can only decay for
smaller scales than that -- i.e.~{\it even if the spectrum were to decay
arbitrary quickly after the USR stage, there would still be a range of
scale-invariant modes after the steep rise.} In the case with an $\eta = -1$ stage presented in 
Fig.~(\ref{1toUSR}d), this range of scales can be reduced by a third.

The impossibility of the spectrum decaying immediately after reaching
the maximum is expected to have an effect on primordial black hole
constraints similar to those derived from the steepest possible growth
by Byrnes {\it et al.}\cite{ByrnesEtAl}, since narrow gaps in PBH
parameter space are more difficult to accommodate. We emphasise that 
the fact that the  $\eta = -1$ phase allows for both a steeper growth 
 and decay alleviates slightly these constraints when compared to the pure $\eta=0$ case. It should be noted, however, that in the presence of a stage of $\eta=-2$ the required duration of the USR stage can be reduced even further while keeping the $k^4$ spectral growth. This is because the decaying mode during the $\eta=-2$ stage decays more slowly than in the $\eta>-2$ cases and thus needs less growth during USR to achieve the same amplitude.

\section{Effect on the spectral index on large scales}
\label{ns}

A further consequence of the generation of steep growth on short scales is a modification
 of the spectral index of larger scales -- even those larger than $k_{\rm dip}$. Since there exist excellent 
constraints on the spectral index, $n_{\rm s}-1 = \di \log {\cal P}_{\cal R}/ \di \log k$, and the running,
$\alpha_{\rm s}=\di n_{\rm s} /  \di \log k$, on CMB scales, one can estimate how far the peak of the
 spectrum has to be from these scales in order not to significantly effect model constraints. 

Considering the slow-roll to USR case we can use Eq.~(\ref{afterSRUSR2}) 
to estimate the contamination in the spectral index 
due to an USR stage. One finds
$\Delta n_{\rm s} \approx -\tau_{USR}^3k^2/k^2_{\rm peak}$. This can be 
compared to the  observational uncertainty in $n_{\rm s}$: 
$\sigma_{n_{\rm s}}=0.0042$~\cite{Akrami:2018odb}. For a USR phase that gives a seven order of 
magnitude growth in the power spectrum, we find modes satisfying
$\log_{10} (k_{\rm peak} /k) \lessapprox 3$ are significantly affected (i.e. $\Delta n_{\rm s}>\sigma_{n_{\rm s}}$).  
Using the matching of the full unexpanded solution we find the more precise condition  $\log_{10} (k_{\rm peak}/k) \lessapprox 3.35$, 
and repeating the calculation for the running, where $\sigma_{\alpha_{\rm s}}=0.0067$, we find $\log_{10} (k_{\rm peak}/k)\lessapprox 3.40$.

\section{Isocurvature modes}
\label{iso}

In a multi-field
scenario, entropy fluctuations may be generated that can then be
converted to curvature perturbations either during or at the end of
inflation. When this conversion takes place, the spectrum in the
isocurvature sector is inherited by the resulting contribution to
${\cal R}$. It is thus conceivable that if isocurvature modes can
have a steeper spectrum than $k^5 (\log k )^2$, a steeper spectrum in
${\cal R}$ could ultimately be achieved.  We give preliminary
consideration to this idea by considering a nearly massless (spectator)
isocurvature mode for which the Sasaki-Mukhanov equation takes the form
\be
v_s''+\left(k^2-a^2H^2(2-\epsilon)\right)v_s=0\,.
\ee
During slow-roll inflation  this equation
implies that the entropy $S\equiv v_s/z$ is conserved on super-horizon scales, with $z$ given by
Eq.~\eqref{Defz2}.  If there is then a transition to a
USR phase or similar, $S$ will grow in a similar way to ${\cal R}$, and
one might think that a matching between the two phases would allow the
scale-dependence of the sub-leading term in an expansion of $S$ before the transition 
to be imprinted on the growing mode after the transition. This cannot occur, however,
 because unlike in the adiabatic case for ${\cal R}$, the
boundary conditions for isocurvature modes imply that $v_s$ and its
first derivative must be continuous, and not the derivative of
$S$. Moreover, because $\eta$ does not enter the Sasaki-Mukhanov
equation for $v_s$, if there is a sharp transition, for example from
slow-roll to USR, its solution is practically
unchanged. It is therefore not possible for the solution after the
transition to capture the scale-dependence of the decaying component
of the pre-transition solution, and the USR phase  simply enhances
the power spectrum with scale-dependence fixed by the leading term
before the transition.

One could also relax the assumption that the isocurvature field is a pure spectator
and allow a fast transition in its mass, as studied in
Ref.~\cite{quench}. A transition which increases the mass parameter
does give rise to a blue spectrum for scales
exiting the horizon after the transition, but one limited by $k^3$.  Moreover, it also induces a
decay of its overall amplitude in time, except in the contrived case 
in which this transition in the
mass is simultaneous with a transition from slow-roll to USR. 

More complicated multi-field scenarios may
generate steeper spectra, but are beyond the scope of this work.

\section{Conclusions}

We have revisited the question of how steeply the power
spectrum from single field inflation can grow with scale. Our first
main result was to show that the ultimate scale-dependence of scales
which exit the horizon during inflation requires us to consider 
the leading and sub-leading terms in an asymptotic expansion of modes normalised 
to the Bunch-Davis vacuum. If there is a 
transition during inflation to a phase with a growing solution (in time) such as USR, the scale-dependence of the sub-leading 
(decaying) mode for scales which exited the horizon before the transition is inherited by the growing mode after the transition.
If USR lasts sufficiently long for the growing mode to become dominant over a range of scales, 
these scales exhibit a steeply growing spectrum.
This insight allowed us to quickly recover the result that a phase of
USR inflation after slow-roll inflation can cause the modes which exit
the horizon towards the end of the slow-roll phase to gain a $k^4$ dependence, and led us 
to our second main result. This was to show that $k^4$ is not 
the steepest sustained growth that is possible, and that a
phase in which $\eta = -1$ followed by a USR phase leads to a $k^5
(\log k)^2$ type growth, which can in principle be sustained 
for an arbitrary large range of scales.

 We also considered over what range of scales the
power spectrum must remain large for after a steep growth and showed 
this can be reduced in cases with the steeper  $k^5
(\log k)^2$ growth. Finally, we
considered over what range of scales $n_{\rm s}$ and $\alpha_{\rm s}$ for
modes which exit during slow-roll are significantly affected by a
subsequent USR phase, and asked whether isocurvature modes could give
rise to an even steeper growth than $k^5 (\log k )^2$, which we found
was likely not possible. We conclude that $k^5 (\log k )^2$ appears to
be the steepest sustained growth of the power spectrum that can be
achieved for inflation with canonical 
scalar fields, and that both this steeper than $k^4$ growth, and subsequent swifter decay, 
could be helpful in avoiding constraints on the formation of PBHs.

\section*{Acknowledgements} 
PC and KAM acknowledge support by STFC grant
ST/P000592/1. DJM is supported by a Royal Society University Research
Fellowship. We thank Philippa Cole and Subodh Patil for discussions.

\appendix

\section{Intermediate cases}\label{Details}

\begin{figure}[]
\centerline{ \hspace{.1cm}\includegraphics[angle=0,width=49mm]{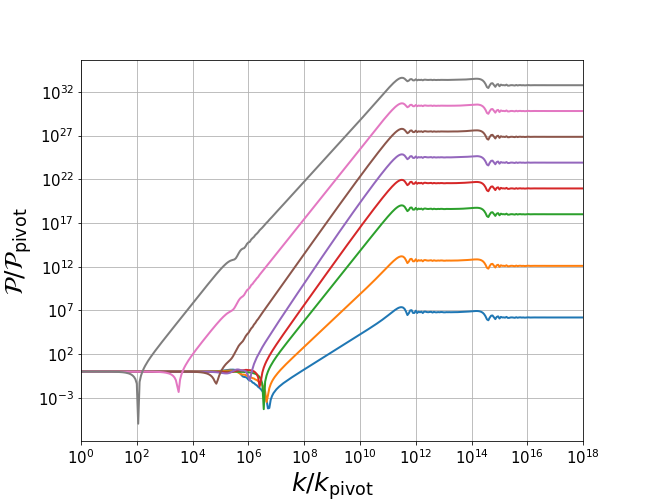} \hspace{-.5cm}\includegraphics[angle=0,width=49mm]{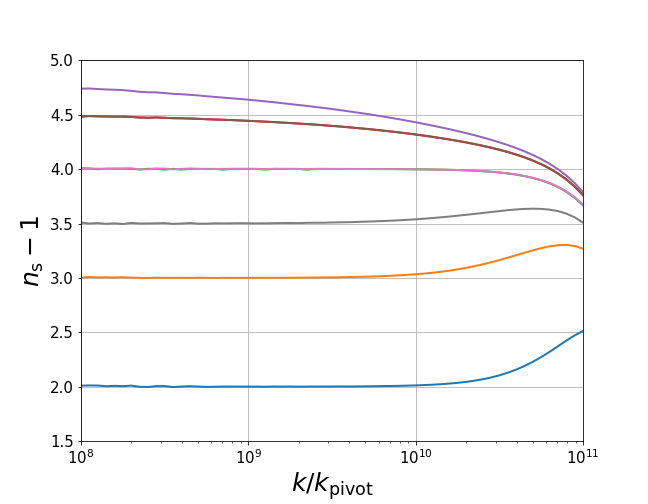}} 
\caption{ Plots of the power spectra (left) and spectral indices (right) for the transitions $\eta=0\rightarrow\eta=\eta_t\rightarrow\eta=-6\rightarrow\eta=0$, with $\eta_t=2,1,0,-1/2,-1,-3/2,-2,-5/2$ (from bottom to top on the left plot). Situations with the same spectral index appear superimposed on the plot on the right, so the lines shown correspond to $\eta_t=2,1,-5/2,-2,-3/2,-1$ (from bottom to top), but include as well $\eta_t=0$ and $\eta_t=-1/2$ in the same positions as $\eta_t=-2$ and $\eta_t=-3/2$, respectively.}
\label{manytoUSR}
\vspace*{-15pt}
\end{figure}

In this appendix, we explore transitions including alternative pre-USR stages not shown in the main text, including scenarios with non-integer values of $\eta$. Our aim is to clearly demonstrate that the spectral index of the decaying term, as shown in Eq.~\eqref{specindfromdecay}, is always imprinted on the growing mode that generates the steepest part of the spectrum, giving a spectral index of $2\alpha+3=5-|\eta+1|$.
We plot our results in Fig.~(\ref{manytoUSR}), in which we can see that the expectations from Eq.~\eqref{specindfromdecay}, hold for almost all cases, including the prediction that the spectral index is the same for situations in which $|\eta+1|$ is the same. For $\eta$ approaching $-1$, the results deviate from a constant spectral index, since higher order terms in a power-series expansion become important. However, the sub-leading term is still a good approximation for scales sufficiently distant from the peak, as can be seen for the cases with $\eta=-1/2$ and $\eta=-3/2$ (superimposed on the right-hand plot). In the $\eta=-1$ limit the power-series expansion breaks down, and many higher order terms effectively contribute and sum to give the logarithmic dependence described in the main text.

\bibliography{mybib}{}

\end{document}